\documentclass[aps, twocolumn, showpacs]{revtex4}
\begin{document}
\title{Topological defects, geometric phases, and the angular momentum of light}
\author { S. C. Tiwari\\
Institute of Natural Philosophy\\
c/o 1 Kusum Kutir Mahamanapuri,Varanasi 221005, India}
\begin{abstract}
 Recent reports on the intriguing features of vector vortex bearing beams are analyzed
using geometric phases in optics. It is argued that the spin redirection phase induced
circular birefringence is the origin of topological phase singularities arising in the
inhomogeneous polarization patterns. A unified picture of recent results is presented based on this proposition. Angular momentum shift within the light beam (OAM)
 has exact equivalence with the angular momentum holonomy associated
with the geometric phase consistent with our conjecture.
\end{abstract}
\maketitle

Topological defects in continuous field theoretic framework are usually associated with
field singularities, however in analogy with crystal defects wavefront dislocations for scalar
waves \cite{1} and disclinations for vector waves \cite{2} have been discussed in the literature.
An important advancement was the realization that topological charge was related with the
orbital angular momentum (OAM) of finite sized (transverse) light beams: typically for the
Laguerre-Gaussian (LG) beams helicoidal spatial structure of the wavefront with azimuthal phase
$exp(i l \phi)$ gives rise to OAM per photon of $l\hbar$ where $l$ is the topological
charge, see review \cite{3}. Adopting the fluid dynamical paradigm topological defects in optics
are termed vortices; singularities in the polarization patterns are called vector vortices
\cite{4}.

The aim of this Letter is to present a unified picture of the underlying physics of intriguing
aspects of recent reports \cite{5,6,7} in terms of the transformation of topological charges
due to  spin redirection phase (SRP) such that OAM is exchanged within the beams \cite{8}.
The role of Pancharatnam phase (PP) invoked in \cite{4,6,7} is also critically examined. 

For the sake of clarity we briefly review the essentials of geometric phases (GP) in optics
which are primarily of two types, see \cite{8,9} for details and original references. Rytov-Vladimirskii phase rediscovered by Chiao and Wu in 1986 (inspired by the Berry phase) arises in the wave vector or momentum space of light. The unit wave vector ${\bf\kappa}$ and polarization vector $\epsilon ({\bf\kappa})$ describe the intrinsic properties of the light wave. A plane wave propagating along a slowly varying path in the real space can be mapped on to
the surface of a unit sphere in the wave vector space, and under  parallel transport along
a curve in this space preserving the spin helicity, ${\bf\epsilon}.{\bf\kappa}$, the
polarization vector is found to be rotated after the completion of a closed cycle on the
sphere. The magnitude of the rotation is given by the solid angle enclosed by the
 cycle, and the sign is determined by the initial polarization state. Since left circular $|L>$
and right circular $|R>$ polarization states acquire equal but opposite geometric phases, Berry terms this effect as geometric circular birefringence \cite{10}.

A polarized light wave propagating in a fixed direction passed through optical elements
traversing a polarization cycle on the Poincare sphere acquires Pancharatnam phase equal 
to half of the solid angle of the cycle. Berry points out that \cite{11} Pancharatnam 
actually made two important contributions. One,a notion of Pancharatnam connection was introduced for the phase difference between two arbitrary nonorthogonal polarizations which 
can be written as $arg({\bf E}{_1}{^*}.{\bf E}_2)$
for complex electric field vectors. Secondly this connection is nontransitive resulting
into the Pancharatnam phase for a geodesic triangle on the sphere. Note that a parallel transport on the Poincare sphere is made with fixed direction of propagation for the occurrence of PP.

In the case of space varying polarization patterns, defining a direction of propagation
is not easy, however Nye \cite{12} has used Pancharatnam connection to define propagation
vector ${\bf k}_\delta$ as a gradient of phase difference between fields at spatial locations
${\bf r}_1$ and ${\bf r}_2$ given by
\begin{equation}
d\delta =Im({\bf E} {^*}.d{\bf E})/|{\bf E}|^2
\end{equation}
In \cite{4} authors correctly use Pancharatnam connection to obtain the phase difference of 
light at two locations in the space varying polarization plane, however the GP involved is not
Pancharatnam phase as one cannot complete polarization cycle without changing the wave vector direction. As discussed above we have to construct appropriate wave vector space for the case of 
vector vortices. For an initial beam propagating along z-axis, at each point $(r,\phi)$ on the inhomogeneous polarization plane there will correspond a ${\bf k}$-space, and spin helicity preserving parallel transport will give
SRP for closed cycles. It is known that for a linearly polarized plane wave
represented by
\begin{equation}
|P>=e^{-i\alpha} |R>+e^{i\alpha} |L>
\end{equation}
 the SRP corresponding to a cycle with solid angle $\Omega$ results into \cite{10}
\begin{equation}
|P{_t}>=e^{-i(\alpha+\Omega)} |R> + e^{i(\alpha+\Omega)} |L>
\end{equation}
For the optical vortices this equation has to be generalized: we suggest a spatially
evolving GP embodied in the solid angle as a function of $(r,\phi)$. This is one of the main contributions of this Letter leading to Eq.(4) below. For the special case in which
only the azimuthal angle dependence matters we obtain $\Omega$ in the
following way. In the transverse plane consider a point A on a circle specified by $\phi$, then
the area enclosed by the great circles in ${\bf k}$-space corresponding to this point and the reference point O specified by $\phi =0$ would subtend a solid angle which varies linearly with 
$\phi$; the solid angle will vary from $0$ to $4\pi$ as $\phi$ varies from $0$ to $2\pi$. Thus we obtain the generalization of Eq.(3) to
\begin{equation}
|P{_v}>=e^{-i(\alpha+2\phi)} |R> + e^{i(\alpha+2\phi} |L>
\end{equation}
Spatially evolving SRP \cite{13} is crucial to understand interesting features observerd in vector 
vortices; we state our second principal contribution in the form of a proposition.

Proposition: Geometric phase induced circular birefringence is the origin of topological
charge transformation in vector vortex carrying beams, and angular momentum
holonomy is manifested as OAM. 

We demonstrate in the following that this proposition offers transparent physical mechanism
to explain the recent reports on inhomogeneous polarization patterns.

Backscattered polarization \cite{5} : Theory and experiments on the backscattered light for linearly polarized light from random media have been of current interest. Interesting features for the backscattering geometry have been observed. Authors of \cite{5} give insightful treatment of the observations invoking GP in wave vector space, and this is in agreement with our proposition. Note that backscattered light wave vector could be treated similar to the discrete transformation
for reflection from a mirror, see discussion in  \cite{9} one can envisage an adiabatic
path in a modified ${\bf k}$-space. It may be noted that essentially
spatially evolving SRP is used in \cite{5}. It seems the term 'geometrical phase vortex'
introduced by them for scalar vortices appearing in space varying polarization pattern is
quite revealing.

The q-plate experiment \cite{6}: An inhomogeneous anisotropic optical element called q-plate 
 has novel addition to HWP : inhomogeneity is introduced orienting the fast (or slow) optical axis making an angle of $\alpha$ with the x-axis in the xy-plane for a planar slab
given by $\alpha(r,\phi) = q \phi + \alpha_0$. Jones calculus
applied at each point of the q-plate shows that the
output beam for an incident $|L>$ state is not only converted to  $|R>$ state but it also acquires an azimuthal phase factor of $exp(2i q\phi)$. Similar to the LG beams this phase is interpreted as OAM of $2q\hbar$ per photon in the output wave. Experiment is carried out using nematic liquid crystal planar cell for $q=1$ and the measurements on the interference pattern formed by the
superposition of the output beam with the reference beam display the wave front 
singularities and helical modes in the output beam.

We argue that in the light of our proposition SRP induced circular birefringence is the origin of topological phase singularity and OAM in q-plates. We picture  helicity preserving transformation  in the  wave vector space defined by ${\bf k}_\delta$. Since the polarization variation is confined in the transverse plane for the q-plates the constraint of the spin helicity
preserving process in the q-plate with the azimuthal
dependence of $\alpha$ would lead to a spiral path for the wave vector.
In q=1 plate the circular plus linear propagation along z-axis will result into a
helical path and the width of the plate ensures that the input and output
ends of the helix are parallel. On the unit sphere in the wave vector space this
will correspond to a great circle, and the solid angle would be $2\pi$. Since the
SRP equals the solid angle for the evolving paths, our Eq.(4) above is in agreement 
with Eq.(3) of \cite{6}. The important observation emphasized by the authors that the incident
polarization controls the sign of the orbital helicity or topological charge is easily
explained in view of the property of the geometric birefingence in which handedness
decides the sign of the phase. Thus both magnitude and sign of the azimuthal phase 
have been explained in accordance with our proposition.

Tightly focused beams \cite{7}: Analysis of the light field at the focal plane of a high
numerical aperture lens for the incoming circularly polarized plane wave shows the existence
of inhomogeneous polarization pattern. Pancharatnam connection at two different points on the
circle around the focus shows $\phi$-dependence of the phase difference. The field can be
decomposed into $|L>$ and $|R>$ states, Eq.(7) in \cite{7}, and it is found that the component
with the spin same as that in the object plane does not change phase while the one with
opposite spin acquires an azimuthal phase of $2\phi$, i. e. topological charge 2 and
OAM of $2\hbar$ per photon. Application of Eq. (4) immediately leads to this result
in conformity with our proposition. We may remark that the construction used by the authors
to derive PP, namely the geodesic triangle on the Poincare sphere formed by the pole,
${\bf E}(r,0)$, and ${\bf E}(r,\phi)$ cannot be completed with a fixed direction of 
propagation for space varying polarization pattern, and therefore it is SRP not PP
that arises.

Having established first part of the proposition, we discuss the role of angular momentum (AM)
holonomy conjecture \cite{8,9}. Transfer of spin AM of light to matter was measured long ago by
Beth \cite{14}, and there are many reports of OAM transfer to particles in recent years \cite{3}.
Since polarization cycle for PP requires spin exchange with optical elements, it is natural to
envisage a role of AM in GP; however it would be trivial. In the AM holonomy conjecture, we argued
AM level shifts within the light beams as physical mechanism for GP. This implies
exact equivalence between AM shift and GP. Indirectly the backscattered light experiments and
their interpretation in terms of SRP supports our conjecture. In the context of the AM conservation \cite{5} the redistribution of total AM within the beam is also indicative
of AM level shifts. The q=1 plate is a special optical element in which no transfer of AM to 
the crystal takes place, and total AM is conserved within the light beam. We argue
that spin is intrinsic, and the OAM is a manifestation of the GP with exact
equivalence between them in this case. The counter-intuitive interpretation in terms of spin to OAM conversion claimed in \cite{6} is clearly ruled out. In fact, Marrucci et al experiment offers first direct evidence in support of our conjecture. It is remarkable that the light field structure calculated for tightly focused beams shows strong resemblence with the action 
of q-plates on light wave, and offers another setting to test our proposition.

To conclude the Letter we make few observations. First let us note that even without the existence
of phase singularities it should be possible to exchange AM within the light beam accompanied
with GP: as argued earlier transverse shifts in the beam would account for the change in OAM 
\cite{9}. Secondly the interplay of evolving GP in space and time domains could be of interest.
 A simple rotating q-plate experiment is suggested: polarized light beam after traversing the q-plate is made to pass through a rotating HWP. Another variant with nonintegral q for this arrangement, i.e. q-plate plus rotating HWP,  is also suggested. Analysis of the emerging beams may delineate the role of SRP and PP as well as provide further test to AM holonomy conjecture. 

The physical mechanism proposed here for  space varying polarization pattern of light could find important application in 'all optical
information processing'. The angular momentum holonomy associated with GP, and the strong evidence
of its proof discussed here will have significance in the context of the controversy surrounding
'the hidden momentum' and Aharanov-Bohm effect. We believe present ideas also hold promise to address some fundamental questions in physics. An important recent example is that of birefringence of the vacuum in quantum electrodynamics in strong external magnetic field. Though
this has been known since long, last year PVLAS experiment reported polarization rotation
\cite{15} apparently very much in excess than the expected one. This has led to a
 controversy on the interpretation of  QED birefringence in external rotating magnetic field, see \cite{16} for a short review. As remarked by Adler
essentially it involves light wave propagation in a nontrivial refracive media, and he finds that to first order there should be no rotation of the polarization of light. Could there be a 
role of GP in this case? It would be interesting to use Pancharatnam connection to calculate the phase of propagating light, and  see if evolving GP in time domain will arise due to rotating magnetic field. It is interesting to note that the magnetic field direction rotates in the  plane transverse to the direction of the propagation of the light. Obviously it would give additional polarization rotation. This problem is being investigated, and will be reported elewhere.

 The Library facility at Banaras Hindu University is acknowledged.


\begin{thebibliography}{99}
\bibitem{1} J. F. Nye and M. V. Berry, Dislocations in wave trains, Proc. R. Soc. Lond. A 336,165 (1974).
\bibitem{2} J. F. Nye, Polarization effects in the diffraction of electromagnetic waves: the role of disclinations, Proc. R. Soc. Lond. A 387, 105 (1983).
\bibitem{3} L. Allen, M. J. Padgett, and M. Babiker, The orbital angular momentum of light, Prog. Opt. 39,291 (1999).
\bibitem{4} A. Niv, G. Biener, V. Kleiner, and E. Hasman, Manipulation of the Pancharatnam phase in vectorial vortices, Opt. Express, 14, 4208 (2006).
\bibitem{5} C. Schwartz and A. Dogariu, Backscattered polarization patterns, optical vortices, and the angular momentum of light, Opt. Lett. 31,1121(2006).
\bibitem{6} L. Marrucci, C. Manzo, and D. Paparo, Optical spin-to-orbital angular momentum conversion in inhomogeneous anisotropic media, Phys. Rev. Lett. 96,163905(2006).
\bibitem{7} Z. Bomzon, M. Gu, and J. Shamir, Angular momentum and geometrical phases in 
tight-focused circularly polarized plane waves, Appl. Phys. Lett. 89,241 (2006).
\bibitem{8} S. C. Tiwari, Geometric phase in optics: quantal or classical?, J. Mod. Opt. 39,1097(1992).
\bibitem{9} S. C. Tiwari, Geometric phase in optics and angular momentum of light, J.Mod. Opt. 51,2297(2004).
\bibitem{10} M. V. Berry, Quantum adiabatic anholonomy, Lectures Ferrara School
on Anomalies, defects, phases..., June 1989.
\bibitem{11} M. V. Berry, The adiabatic phase and Pancharatnam's phase for polarized light, J. Mod. Opt. 34, 1401 (1987).
\bibitem{12} J. F. Nye, Phase gradient and crystal-like geometry in electromagnetic and elastic wavefields, in Sir Charles Frank OBE, FRS:An eightieth birthday tribute 
(IOP,UK 1991)pp220-231.
\bibitem{13} S. C. Tiwari, Nature of the angular momentum of light: rotational energy and geometric phase, arxiv.org : quant-ph/0609015.
\bibitem{14} R. A. Beth, Direct detection of the angular momentum of light, Phys. Rev. 48, 471 (1935).
\bibitem{15} E. Zavattini et al, Experimental observation of optical rotation generated in vacuum by a magnetic field, Phys. Rev. Lett. 96, 110406 (2006).
\bibitem{16} S. L. Adler, Vacuum birefringence in a rotating magnetic field, J. Phys. A: Math. Theor. 40, F143 (2007).

\end{thebibliography}
\end{document}